\documentstyle[pra,preprint,aps]{revtex}
\renewcommand{\baselinestretch}{1}
\begin{document}
\title{Exact remote state preparation for multiparties }
\author{P. Agrawal$^{(1)}$, P. Parashar$^{(2)}$, and A. K. Pati$^{(1,3)}$ }
\address{$^{(1)}$Institute of Physics, Sainik School Post, 
Bhubaneswar-751005, India} 
\address{$^{(2)}$ Physics and Applied Mathematics Unit, Indian Statistical 
Institute, Kolkata, India}
\address{$^{(3)}$School of Informatics, University of Wales, Bangor LL 57 
1UT, UK}

\maketitle
\def\ra{\rangle}
\def\la{\langle}
\def\ver{\arrowvert}

\begin{abstract}
We discuss the exact remote state preparation protocol of special 
ensembles of qubits at multiple locations. We also present generalization 
of this protocol for higher dimensional 
Hilbert space systems for
multiparties. Using the `dark states', the analogue of singlet EPR pair 
for multiparties in higher dimension as quantum channel, we show several
instances of remote state preparation protocol using multiparticle measurement
and classical communication. 

\end{abstract}

\vskip .5cm

PACS           NO:    03.67.-a, 03.65.Bz\\

Email: agrawal@iopb.res.in, parashar@isical.ac.in, akpati@iopb.res.in

\vskip 1cm

\par

\section{Introduction}
Quantum information theory has opened up the possibility of novel form
of information processing tasks which are not possible classically.
Two important quantum information processing tasks in recent
years have been teleportation \cite{cb} and remote state preparation (RSP)
\cite{akp}.
In quantum teleportation the sender and the receiver 
do not know the identity of a state. In remote state preparation the sender 
wants to prepare a state of her choice at a distant lab, thus she knows the 
state which is to be remotely prepared. It was found that special class of 
qubits can be remotely prepared using one unit of entanglement and one 
classical bit \cite{akp}.
Furthermore, if the aim is not to prepare an arbitrary qubit, rather to
simulate the measurement statistics on a qubit, then it is possible for
Alice to do that at Bob's place with one ebit and one cbit. This is called
remote state measurement protocol (RSM) \cite{akp}.
There has been considerable interest in preparation of quantum states
at a remote location using previously shared entanglement, local operation
and classical communication 
\cite{akp,hklo,betal,db,zeng,debi,berry,peng,peng1}.

Unlike in teleportation where the resources are fixed for the task, here it
is possible to have trade-offs. It was  conjectured by Lo \cite{hklo}
that if Alice wants to prepare remotely an arbitrary qubit it may still 
require two classical bits as in the case of quantum teleportation. 
Bennett {\it et al} have generalized RSP for arbitrary qubits, higher 
dimensional Hilbert spaces and also of entangled systems \cite{betal}.
Devetak and Berger have proposed a low entanglement RSP protocol \cite{db} 
for arbitrary quantum states. 
The exact and minimal resource consuming RSP protocol
is generalized to higher dimension by  Zeng and Zhang \cite{zeng}.
There are restrictions on the dimension of
the Hilbert space for which RSP can be realized. 
Leung and Shor have given a stronger proof of Lo's conjecture for RSP of
arbitrary quantum state \cite{debi}. Remote preparation of ensemble of
mixed states has been studied by Berry and Sanders \cite{berry}. 
The exact RSP and RSM protocols for qubits \cite{akp} have been implemented 
using NMR devices \cite{peng,peng1} over inter-atomic distances. 
Also, there has been a recent attempt to generalise RSP of a equatorial qubit
at two locations in an approximate manner \cite{yfz}.

In this paper we would like to generalize RSP protocol for multiparties in an
exact manner. The
task here is the following: How can Alice prepare a quantum state of her 
choice at various
locations (say at Bob, Charlie, Denis,.. and so on) using previously shared
entanglement, local operation and classical communication ? Unlike in 
teleportation, we are allowed to ask this question in RSP. In quantum 
teleportation one can create a replica of a quantum state at one place only
at the expense of destroying the original, so as not to violate the no-cloning
principle \cite{wz,dd}. In RSP Alice knows the state, hence she can prepare
as many copies as she wants. Of course preparing copies in her lab does not
need any entanglement or classical communication but as expected 
in a distant lab she does need quantum and classical resources to accomplish
the task. That is the subject of the present paper.
In order to be able to perform RSP at multiple locations,
the first question is: what kind of quantum resource does one need
between multiusers? We find that the so called  `dark states'\cite{kok} play a
crucial role. The general finding here is that Alice can prepare a known state
of special class of qudit at multiple locations by performing multiparticle 
measurement and sending $\log_2 d$ cbits of information to each party. 
The organization of the paper is as 
follows. In  section II, we discuss quantum 
resource suitable for exact RSP for multiparties. In section III, we
present a protocol for RSP of special class of qubits for
multiparties. We also provide a 
probabilistic RSP protocol for qubits. Section IV is devoted to the
generalization of the
scheme for special class of qutrits. In section V, we extend 
our protocol to  
higher dimensional systems,  i.e., qudits. Finally, we close the
paper with some conclusions.

\section{Quantum resource}

It is worth observing that in the case of exact RSP protocol of special 
ensemble of qubits one uses  singlet EPR (Einstein-Podolsky-Rosen) 
pairs $\ver \Psi^- \ra = 
{1 \over \sqrt 2}(\ver 0 \ra \ver 1 \ra - \ver 1 \ra \ver 0 \ra)$
which are invariant under local unitary transformations, i.e., 
$U \otimes U \ver \Psi^- \ra = 
\ver \Psi^- \ra$. Another remarkable property follows from the above 
invariance nature of EPR singlet is that if a subsystem undergoes an 
evolution, then the other subsystem undergoes a reverse-evolution 
or vice versa. This is really counter intuitive that has no classical 
analog  \cite{akp1}! It is expressed by the following equation:
\begin{eqnarray}
U^{\dagger} \otimes I \ver \Psi^- \ra = 
I \otimes U\ver \Psi^- \ra.
\end{eqnarray}
However, this evolution cannot be seen at individual level, because the 
state of either qubit is described by a completely random density matrix. 
The evolution leading to 
state preparation is possible only after local measurement and 
sending the classical information. This property is very crucial, because,
in some sense such a state has all possible information (complement 
information) about a qubit. Therefore, it is legitimate
to look for such states for multiparties that enjoy the above invariance
property. Suppose we have a composite system with each subsystem being 
in a Hilbert space of dimension $d$. An $N$-particle entangled state 
$\ver \Psi \ra$ with the property $U \otimes U \otimes \cdots \otimes 
U \ver \Psi \ra = 
\ver \Psi \ra$ would be a useful quantum resource for multiparty RSP. 
Such states are called {\em dark states}. Essentially, they live 
in a `robust part'
of the Hilbert space where if all the particles are subjected to same 
unitaries then nothing happens, but if one particle is 
subjected to a unitary operator then that is equivalent to applying inverse
unitary transformation to rest of the particles. This can
be seen from the following equation 
\begin{eqnarray}
U^{\dagger} \otimes I \otimes I \otimes \cdots \otimes I \ver \Psi \ra = 
I \otimes U \otimes U \cdots \otimes U \ver \Psi \ra.
\end{eqnarray}
 Here, we briefly recapitulate the essential
properties of dark states from Ref.\cite{kok} which are useful for
our purpose. These states are 
the eigenstates of the interaction Hamiltonian with eigenvalue zero and 
hence do not evolve in time. 
There are no bipartite dark states for the systems with Hilbert space
of dimension $d > 2$. The smallest 
system of qudits in a dark state is a $d$-partite quantum system. 
In general, dark states exist for a d-level N-particle system
only if $N= md$, with $m$ being the set of natural numbers.
Also, coherent and incoherent superposition of dark states is also
 a dark state, i.e, if $|\Psi\ra$ and $|\Phi\ra$ are two dark states, then
 $a |\Psi\ra + b |\Phi\ra$ and $p |\Psi\ra \la \Psi| + q |\Phi\ra \la \Phi|$
 are also dark states. However, we wish to emphasize that for a given system
there are dark states which are not useful for remote state preparation,
as we shall also see below.

\section{Remote State Preparation of a Qubit}

An arbitrary state of a qubit can be represented as,

\begin{equation}
    |\psi\ra = cos(\theta/2) |0\ra + \,  sin(\theta/2)
                             e^{i \varphi } |1\ra
\end{equation}
       
       Here $\{ |0\ra, |1\ra\}$ are called computational basis vectors.
     There are two real parameters $ 0 \leq \theta  \leq \pi$ and 
     $ 0 \leq \varphi  \leq 2 \pi$. The angles $\theta$ and $\varphi$
     define a point on the unit two-dimensional sphere, known as  
        {\it Bloch sphere}.
  It also corresponds to the state of a spin-$1 \over 2$ particle
    (up to an overall phase)
    with the direction of the spin specified by  $\theta$ and $\varphi$.
  The Hilbert space of a qubit is two
        dimensional, so one can also choose the basis vectors as 
     $\{ |\psi\ra,  |{\bar \psi}\ra \}$, such that the state
$ |{\bar \psi}\ra \}$ is orthogonal to the state  $|\psi\ra$,
and  $|{\bar \psi}\ra \}$ is given by
\begin{equation}
     |{\bar \psi} \ra = - sin(\theta/2) |0\ra + \, cos(\theta/2)
                                           e^{i \varphi } |1\ra.
\end{equation}

\subsection{Exact remote state preparation for one party}

       The remote state preparation protocol for a special class of
       qubit states was introduced in Ref \cite{akp}. Here we review
       this protocol.
       Alice and Bob share one qubit each which are in an entangled
       state. As discussed above, this entangled state has to be a
       dark state for the success of the protocol. In the case of
       two parties (Alice and Bob) and two-dimensional Hilbert
       spaces of qubits, such a state is the singlet state: 
\begin{equation}
    |\Psi^{-}\ra = {1\over \sqrt{2}}( |0\ra|1\ra - |1\ra|0\ra).
\end{equation}

      Alice now wishes to help Bob to prepare a state that is known
     completely only to her. Bob may also know the value of one of
     the parameters; so he knows the ensemble to which the state 
     corresponds to. According to the protocol, Alice first applies
     an unitary transformation on her qubit. This unitary
     transformation changes  $\{ |0\ra,  |1\ra \}$ to
       $\{ |\psi\ra,  |{\bar \psi}\ra \}$, where $|\psi\ra$ is the state
     that Alice wishes Bob to prepare. To illustrate the protocol,
     let us first consider the following ensemble of states, 
 
\begin{equation}
    |\psi\ra = cos(\theta/2) |0\ra + \, sin(\theta/2) |1\ra.
\end{equation}

       For this ensemble of states, $\varphi = 0$. These states 
    belong to polar great circle on the Bloch sphere.
    Alice performs a unitary
    transformation, determined by the angle $\theta$, on her qubit.
    As discussed earlier, this can correspond to the following change
    in the shared entangled state:
\begin{equation}
I\otimes U(\theta)|\Psi^-\ra = {1\over \sqrt{2}}( |0 \ra|{\bar \psi}\ra - \, |1\ra|\psi\ra).
\end{equation}

    Here $|{\bar \psi}\ra$ is as given in equation $(4)$ with $\varphi = 0$.
    After making the transformation Alice makes a measurement on her
    qubit using the basis vectors  $\{ |0\ra,  |1\ra \}$. Then the state
    of Bob's qubit can be either $|\psi\ra$ or $ |{\bar \psi}\ra$.
    If the state is  $ |{\bar \psi}\ra$, Bob can convert it to the 
    desired state  $|\psi\ra$ by a rotation by $\pi$
    around $y-$axis. The rotation operator is $i \sigma_{y}$.
    After making the measurement, Alice sends Bob one cbit of
    information, leading Bob to do either nothing or apply $i \sigma_{y}$.

     There are other ensembles of states, that can be remotely prepared
    using the above protocol. Let us discuss these ensembles.
The one such ensemble corresponds to the equatorial qubit states.
    For such states, $\theta = \pi/2$ and we have:

\begin{equation}
    |\psi\ra ={1\over \sqrt{2}} ( |0\ra + \, e^{i \varphi} |1\ra).
\end{equation}
Here Bob can obtain $ |\psi\ra$ from $|{\bar \psi}\ra$ up to a phase
    by a rotation by $\pi$ around $z-$axis, i.e. by
    applying $\sigma_{z}$ to $|{\bar \psi}\ra$.
Another example of the  ensemble of the states is
 given by $\varphi = \pi/2$. This is another class
 of polar qubit states:

\begin{equation}
    |\psi\ra = cos(\theta/2) |0\ra + \, i sin(\theta/2)) |1\ra
\end{equation}
Here Bob can obtain $ |\psi\ra$ from $|{\bar \psi}\ra$ up to a phase
    by a rotation by $\pi$ around $x-$axis, i.e., by
    applying $\sigma_{x}$ to the $|{\bar \psi}\ra$.

    In the above, we have considered ensembles of states with two 
   different values of the parameter $\varphi$. In fact the protocol
   works for any choice of $\varphi$. Suppose the state
   Alice wishes to remotely prepare belongs to the ensemble of states
   with $\varphi = \varphi_{0}$ \cite{peng1}:
   
\begin{equation}
    |\psi\ra = cos(\theta/2) |0\ra + \, sin(\theta/2))
                             e^{i \varphi_{0} } |1\ra
\end{equation}
Bob can obtain $ |\psi\ra$ from $ |{\bar \psi}\ra$ up to a phase
by applying the following unitary operator:
$$
\left(\matrix{ 0 &  e^{-i \varphi_{0}} \cr - e^{i \varphi_{0}} & 0} \right).
$$

Instead of fixing the parameter $\varphi$ in $(3)$, we may 
fix the parameter $\theta = \theta_{0}$ and leave $\varphi$
arbitrary. A state from such an ensemble can be remotely prepared
if we can find unitary transformations connecting 
$ |\psi\ra$ and $ |{\bar \psi}\ra$ which are independent of
$\varphi$. It turns out that connecting transformations
are Hermitian not unitary \cite{herm}.

\subsection{Exact remote state preparation of qubits for multiparties}

 In this section, we wish to explore the possibility of Alice
helping many parties to remotely prepare identical states by making
a single measurement on her qubits. Suppose there are $N$ parties.
Then the minimum number of qubits they can share with Alice is $N+1$.
But these qubits cannot necessarily be in a dark state.
As discussed in section II, for N-partite  entangled 
qubit dark states condition $N=2m$ holds. Therefore, for 
only even number of qubits, there can be a dark state.
It turns out that if Alice makes a one-particle measurement then
exact remote state preparation by many parties is not possible. In
such a case, the state of the qubits belonging to different
parties remain entangled.
However, if Alice makes a multiparticle measurement, then the protocol
as discussed for the one party works also for multiparties.
Here, we explicitly give a protocol for RSP of special ensemble of qubits
at two locations {\em simultaneously}. Let us suppose that
 Alice  chooses to prepare 
a qubit from the class given in equation $(6)$.

   To {\em simultaneously} prepare a state at two locations, 
  we need four qubits to use a dark state as a resource. Alice
  has two qubits. Other two parties, Bob and Charlie,
  have one qubit each. The quantum resource here would be 
  the four-qubit dark state, 
\begin{eqnarray}
\ver \Psi \ra_{1234} = 
{1 \over  2}[ \ver 0011 \ra +  \ver 1100 \ra 
- \ver 0110 \ra - \ver 1001 \ra ] .
\end{eqnarray}
Let Alice possess particles $(1,2)$; Bob has particle $3$ and Charlie has
particle $4$. 
Alice applies local unitary transformations $U^{\dagger} \otimes U^{\dagger}$ 
to her qubits that brings the above state to
\begin{eqnarray}
U^{\dagger} \otimes U^{\dagger} \otimes I  \otimes I \ver \Psi \ra &=& 
I \otimes I \otimes U \otimes U \ver \Psi \ra \nonumber\\
&=& {1 \over  2}[ \ver 00 \ra \ver {\bar \psi} \ra \ver{\bar \psi} \ra +  
\ver 11 \ra \ver \psi \ra \ver \psi \ra 
- \ver 01 \ra \ver {\bar \psi} \ra \ver \psi \ra - 
\ver 10 \ra \ver \psi \ra \ver {\bar \psi} \ra ] 
\end{eqnarray}

Here $\ver \psi \ra$ and $ \ver {\bar \psi} \ra$ are as given in
equations $(3)$ and $(4)$ with $\varphi = 0$.
Alice carries out a von Neumann projection onto 
two qubit basis  $\{\ver 00 \ra, \ver 01\ra, \ver 10 \ra, 
\ver 11 \ra \}$ and sends one classical bit each to Bob and Charlie. 
The state of Bob's qubit can be either $\ver \psi \ra$ or 
$ \ver {\bar \psi} \ra$; the same is true for Charlie. Alice
has to convey to them by sending one cbit each whether to
apply the operator  $ i \sigma_y$ on their qubit or not.
As we discussed in the last section, apart from the polar qubit
of equation $(6)$, equatorial qubit of equation (8) and polar qubits
of equations $(9)$ and $(10)$ can also be remotely prepared by one
party. Following the above protocol, Alice can help prepare
these ensembles of qubits at two locations. The only difference
will be that Bob and Charlie would apply the unitary operator
$\sigma_z$, or $\sigma_x$, or the given after equation $(10)$, 
depending on the ensemble.
This will constitute RSP of these special ensemble of qubits 
simultaneously for two parties. The amount of quantum and 
classical resources used here is two ebits and two cbits. 

Apart from the dark state of equation $(11)$, we can also use the following
dark state for the remote state preparation at two locations,
\begin{eqnarray}
\ver \Psi_{1} \ra_{1234} = 
{1 \over  2}[ \ver 0011 \ra +  \ver 1100 \ra 
- \ver 0101 \ra - \ver 1010 \ra ] .
\end{eqnarray}

Other dark states, some of which can be obtained from
the linear combination of the above two dark states, would {\it not}
help in remote state preparation. In these cases, the state of 
Bob and Charlie's qubits will remain entangled even after Alice
has made measurement on her two qubits. The key property of the dark
states of equations $(11)$ and $(13)$ is that they can be written as
$\ver \Psi^{-} \ra_{13} \otimes \ver \Psi^{-} \ra_{24}$
and $\ver \Psi^{-} \ra_{14} \otimes \ver \Psi^{-} \ra_{23}$
respectively. (Here $\ver \Psi^{-} \ra_{ij}$ is a EPR singlet state
as given in equation (5).) When Alice makes a measurement on her
qubits $(1,2)$, then the qubits of Bob and Charlie are no longer
entangled.

Note that we are not repeating the exact RSP protocol \cite{akp} in sequence.
Even though virtually we use same number of EPR pairs (two ebits), we can 
perform RSP of a qubit from these special ensemble at two locations in a 
{\em single shot}. Here Alice is making a multiparticle measurement; not 
one-particle measurements in sequence.

In order to prepare a qubit at $m$ locations, we can follow the same protocol
as above. We start with an entangled state consisting of $N=2m$ qubits, 
of which $m$ qubits will be at Alice's 
location and the rest $m$ qubits are located with $m$ parties at different
locations. There are many dark states which can be used as a quantum
resource. (We conjecture that there are $m!$ such resource states.)
One such shared resource state can be explicitly given as,
\begin{eqnarray}
\ver \Psi \ra_{12\ldots 2m} &=& \frac{1}{2^m} 
(\ver 0 \ra_1 |1\ra_{m+1} - |1\ra_1 |0\ra_{m+1}) \otimes
(\ver 0 \ra_2 |1\ra_{m+2} - |1\ra_2 |0\ra_{m+2}) \nonumber\\
&\otimes& \cdots \otimes
(\ver 0 \ra_m |1\ra_{2m} - |1\ra_m |0\ra_{2m}).
\end{eqnarray}

  Alice can help prepare remotely any of the states from the 
ensembles discussed in the last section. Let $\ver \psi \ra $
be one such state. Alice makes unitary transformations on her
$m$ qubits so that $\{ \ver 0 \ra, \ver 1 \ra \} \to
\{ \ver \psi \ra, \ver {\bar \psi} \ra \}$ at remote locations.
Next, Alice projects onto her $m$ qubits. After the measurement,
Alice can send one cbit each
to $m$ parties so that they can apply appropriate unitary
transformation on their qubit to prepare the state $\ver \psi \ra $.
Thus she can prepare a qubit from these special ensembles at $m$ 
locations using only $m$-ebits and $m$-cbits.

\subsection{Probabilistic remote state preparation of a qubit}

Exact remote preparation is not possible with all dark states.
However, there is a finite probability for remote state 
preparation with any dark state.  Let us illustrate it with
the following example. For four qubit case, 
two of the dark states are 
given by $\ver \Psi^- \ra_{12} \otimes \ver \Psi^- \ra_{34}$ 
and $\ver \Psi^- \ra_{13} \otimes \ver \Psi^- \ra_{24}$. 
The former is not useful for RSP due to the presence of local entanglement
between particles $1$ and $2$, and similarly between $3$ and $4$. 
 The latter one is the state $(11)$ used in
 the exact RSP protocol. Since any linear
 superposition of two dark states is also a dark state, let us consider a
general superposition of these two states
\begin{equation}
\ver \Phi \ra_{1234}= N[a \ver \Psi^- \ra_{13} \otimes \ver \Psi^- \ra_{24} + 
b \ver \Psi^- \ra_{12} \otimes \ver \Psi^- \ra_{34}],
\end{equation}
which can be written as
\begin{eqnarray}
\ver \Phi \ra_{1234} &=& N[a \ver 0011 \ra_{1234} + a \ver 1100 \ra_{1234} -
(a+b) \ver 0110 \ra_{1234} \nonumber\\
&-& (a+b) \ver 1001 \ra_{1234}
+ b \ver 0101 \ra_{1234} +b \ver 1010 \ra_{1234}],
\end{eqnarray}
where $N= 1/2\sqrt{a^2 +b^2 + ab}$ is the normalization constant.

Let the resource shared between Alice, Bob and Charlie be given by
the above entangled state. Alice has particles $(1,2)$, Bob has particle
$3$ and Charlie has particle $4$. Alice applies a unitary operator to her
particles that brings the above state to

\begin{eqnarray}
U^{\dagger} \otimes U^{\dagger} \otimes I  \otimes I \ver \Phi \ra &=& 
N[a \ver 00 \ra_{12} \ver {\bar \psi}{\bar \psi} \ra_{34} + a \ver 11 \ra_{12}
\ver \psi \psi\ra_{34} - (a+b) \ver 01 \ra_{12} 
\ver {\bar \psi} \psi \ra_{34} \nonumber\\
&-& (a+b) \ver 10 \ra_{12} \ver \psi {\bar \psi} \ra_{34}
+ b \ver 01 \ra_{12} \ver \psi {\bar \psi} \ra_{34} + 
b \ver 10 \ra_{12} \ver {\bar \psi} \psi \ra_{34}].
\end{eqnarray}

Here  $\ver \psi \ra$ is the state belonging to one of the ensembles
discussed above.
Now, Alice carries out projection measurement onto two qubit basis
$\{\ver 00 \ra, \ver 01\ra, \ver 10 \ra, 
\ver 11 \ra \}$ and sends one classical bit each to Bob and Charlie. 
If her outcome is $\ver 00 \ra$, then 
Bob and Charlie's qubits would be in the state $\ver {\bar \psi} \ra$.
Therefore, after receiving classical communication each of them has to
apply appropriate unitary transformation
to correct the state. The probability of this occurrence is $P_{00}=
a^2/{4(a^2 +b^2 + ab)}$.
If she gets $\ver 11 \ra$, then Bob and Charlie need not do anything. The 
probability of occurrence is $P_{11}= a^2/{4(a^2 +b^2 + ab)}$. However, 
if she gets $\ver 01 \ra$, then the particles at Bob and Charlie's locations
are in an entangled state given by 
\begin{eqnarray}
(a+b) \ver {\bar \psi} \ra_{3} \ver \psi \ra_{4}
+ b \ver \psi \ra_{3} \ver {\bar \psi} \ra_{4}.
\end{eqnarray}
This occurs with probability $P_{01}= ((a+b)^2 + b^2)/{4(a^2 +b^2 + ab)}$.
Since the qubit state is unknown to Bob and Charlie both, they cannot
disentangle it and apply some local unitary operation to get the original 
state. Hence, this event can be regarded as the failure one. Similarly, when
Alice gets $\ver 10 \ra$, then Bob and Charlie's qubits are in an entangled
state
\begin{eqnarray}
(a+b) \ver \psi \ra_{3} \ver {\bar \psi} \ra_{4}
+ b \ver {\bar \psi} \ra_{3} \ver \psi \ra_{4}.
\end{eqnarray}
This also occurs with probability $P_{10}= ((a+b)^2 + b^2)/{4(a^2 +b^2 + ab)}$
and as before, they cannot disentangle it exactly. So this protocol is
probabilistic with a success probability given by
\begin{eqnarray}
P_S= \frac{a^2}{2(a^2 +b^2 + ab)}.
\end{eqnarray}

We note that one classical bit of information would be enough
if Bob and Charlie need to have communication with Alice only
in the event of success of the protocol. In case, Alice has to 
communicate the failure also, then she needs to send $\log_{2}3$
to each party. If Bob and Charlie wish to cooperate and do some joint action
to recover the state of a qubit, then Alice needs to communicate 
two classical bits (i.e. all four possible outcomes) to each party.

Thus with an arbitrary superposition of dark states one can have probabilistic
remote state preparation of a qubit at multiple location. 
The amount of entanglement between particles $(1,2)$ versus $(3,4)$
is 
$E(\Phi)= -3N^2a^2\log_2 N^2a^2 - N^2(a + 2b)^2 \log_2 N^2(a + 2b)^2$
which is less than two ebits.
 
Because there is a component from the
so-called useless resource (the local entanglement between qubits $1$ and
$2$, and similarly between $3$ and $4$) we have a probability of failure. 
This brings out another
feature of quantum communication: the presence of local
entanglement which is thought of as not `good', in fact 
plays a bad role, in 
the sense that its superposition with the `shared resource' part can sometimes 
lead to failure of the protocol.

\section{Remote State Preparation of a Qutrit}

We now turn our attention to the case of a qutrit, where 
the dimension of the Hilbert space is three.
A qutrit $|\psi\ra \in {\cal H}^3$ can be parametrized by four real 
parameters $\gamma_1, \gamma_2,
\delta $ and $\phi$  such that $0 \leq \gamma_1, \gamma_2 \leq \pi/2$ and 
$0 \leq \delta, \phi \leq 2\pi$. The most general qutrit state can be
expressed as
\begin{eqnarray}
\ver \psi \ra = cos \gamma_1 \ver 0 \ra + sin \gamma_1 cos \gamma_2 
e^{i\delta} \ver 1 \ra + sin \gamma_1 sin \gamma_2 e^{i\phi} \ver 2 \ra
\end{eqnarray}

Ideally, Alice's aim is to prepare this most general state remotely. 
But as in qubit case, such a state cannot be remotely prepared using
exact RSP protocol. With this protocol, it can be prepared remotely only 
probabilistically.
However, as we shall see below, just as in the case of qubits,
there exist ensembles of states where exact RSP can be performed
with multiparticle measurements. 

For a qutrit, there exist many sets of basis vectors which include
the state $(21)$. One such set can be obtained by applying
a specific unitary transformation on the computational basis vectors, 
\begin{eqnarray}
U(\gamma_1,\gamma_2,\delta,\phi) \ver 0 \ra &=& \ver \psi_0 \ra = 
cos \gamma_1 \ver 0 \ra + sin \gamma_1
cos \gamma_2 e^{i\delta} \ver 1 \ra + sin \gamma_1 sin \gamma_2 e^{i\phi}
\ver 2 \ra \nonumber\\
U(\gamma_1,\gamma_2,\delta,\phi) \ver 1 \ra &=& \ver \psi_1 \ra 
= sin \gamma_1 \ver 0 \ra - cos \gamma_1 
cos \gamma_2 e^{i\delta} \ver 1 \ra - cos \gamma_1 sin \gamma_2 e^{i\phi} 
\ver 2 \ra  \nonumber\\
U(\gamma_1,\gamma_2,\delta,\phi) \ver 2 \ra &=& \ver \psi_2 \ra 
= sin \gamma_2 e^{i\delta} \ver 1 \ra
- cos \gamma_2 e^{i\phi} \ver 2 \ra 
\end{eqnarray}

 The states $\ver \psi_0 \ra, \ver \psi_1 \ra$ and $\ver \psi_2 \ra$
are orthogonal to each other.

\subsection{ Exact remote state preparation for one party}

Suppose Alice wishes to prepare a qutrit state at Bob's location.
For this purpose, she would need 
to share a dark state $\ver \Phi \ra$ with Bob. If Alice and
Bob have one qutrit each, then these qutrits cannot be in a
dark state. As discussed in section II, a
minimum of three qutrits are needed to construct a
dark state. If Alice has one qutrit, then Bob would need to
have two qutrits. In this case, 
Alice's measurement would leave Bob's two qutrits in an entangled state.
Then remote state preparation would not be possible. So, Alice 
may have two qutrits, while Bob would have one.  

Dark state coincides with the antisymmetric state when the number 
of particles $N$ equals the dimension of the Hilbert space $d$ of each 
particle. (For example, when $N=d=2$ we get the singlet state.)  
Let $\{ \ver 0 \ra , \ver 1 \ra , \ver 2 \ra \}$ be the orthonormal basis of 
the qutrit. Then the shared resource between Alice and Bob is
\begin{eqnarray}
\ver \Phi \ra_{123} = {1 \over \sqrt 6} 
[ \ver 012 \ra + \ver 120 \ra + \ver 201 
\ra - \ver 021 \ra - \ver 102 \ra - \ver 210 \ra ]  
\end{eqnarray}

Since the resource state $(23)$ is a dark state, the action of the 
unitary operator $U\otimes U \otimes U$ on it leaves it invariant. 
We suppose that Alice possesses particles $(1, 2)$ 
while particle 3 is with Bob.
Now Alice applies $U^{\dagger} \otimes U^{\dagger}$ to her particles and
as a result we have
\begin{eqnarray}
U^{\dagger} \otimes U^{\dagger} \otimes I \ver \Phi \ra &=& {1 \over \sqrt 6}
[\ver 0 1\ra \ver \psi_2 \ra + \ver 1 2 \ra 
\ver \psi_0 \ra + \ver 2 0 \ra \ver \psi_1 \ra \nonumber \\
&-& \ver 0 2 \ra \ver \psi_1 \ra - \ver 1 0 \ra 
\psi_2 \ra - \ver 2 1\ra \ver \psi_0 \ra ].
\end{eqnarray}

Here $\ver \psi_0 \ra, \ver \psi_1 \ra$ and $ \ver \psi_2 \ra$
form a set of basis vectors for the qutrit.
If Alice carries out a two-qutrit orthogonal measurement,
Bob's state would always be in one of the three basis states.
As in the qubit case, in general, one cannot find parameter
independent unitary transformations to change one basis
vector to another. It is possible for only some ensembles
of states. Below we discuss one such ensemble of states.

Let us now consider the following qutrit state  
which Alice wishes to prepare remotely, 
\begin{eqnarray}
\ver \psi \ra = { 1\over \sqrt 3} ( \ver 0 \ra + 
e^{i\delta} \ver 1 \ra + e^{i\phi} \ver 2 \ra )
\end{eqnarray}
where $\delta, \phi$ are arbitrary. Such a state resembles in 
form the ``equatorial'' qubit state. This belongs to
a specific ensemble with $\gamma_2 = \pi/4$ and $\gamma_1$
such that $cos\gamma_{1} = 1/\sqrt{3}$ in the qutrit
state $(21)$; phases are arbitrary (known to Alice but unknown to Bob).

If one applies a unitary transformation on the computational basis vectors 
then one obtains
\begin{eqnarray}
U \ver 0 \ra = \ver \psi_0 \ra = {1 \over \sqrt 3} ( \ver 0 \ra +
e^{i\delta} \ver 1 \ra
+ e^{i\phi} \ver 2 \ra)
\end{eqnarray}
\begin{eqnarray}
U \ver 1 \ra = \ver \psi_1 \ra = {1 \over \sqrt 3} ( \ver 0
\ra + \Gamma
 e^{i\delta} \ver 1 \ra + {\Gamma}^2 e^{i\phi}
\ver 2 \ra )
\end{eqnarray}
\begin{eqnarray}
U \ver 2 \ra = \ver \psi_2 \ra = {1 \over \sqrt 3} ( \ver 0
\ra + {\Gamma}^2
 e^{i\delta} \ver 1 \ra +  \Gamma  e^{i\phi}
\ver 2 \ra )
\end{eqnarray}
where $\Gamma = e^{2\pi i/3}$. The set  
 $\{\ver \psi_0 \ra, \ver \psi_1 \ra, \ver \psi_3
\ra \}$ form an orthonormal basis \cite{cerf} and is related to the     
computational basis by the discrete Fourier Transform.
Here $\ver \psi_0 \ra$ is the state that Alice wishes to prepare
remotely. $\ver \psi_1 \ra$ can be transformed into $\ver \psi_0 \ra$
by the unitary transformation $U_{01} = diag ( 1, {\Gamma}^{2}, \Gamma )$.
Similarly, $\ver \psi_2 \ra$ can be transformed into $\ver \psi_0 \ra$
by the unitary transformation $U_{02} = diag ( 1, \Gamma, \Gamma^{2} )$.

Let us now carry out the protocol.
Alice first applies appropriate unitary transformation to transform
the dark state as in equation $(23)$. Next, she makes a two-particle
measurement in the basis $\{\ver 0 0\ra, \ver 0 1\ra, \ver 0 2 \ra, 
\ver 1 0 \ra, \ver 1 1\ra, \ver 1 2 \ra, \ver 2 0 \ra, \ver 2 1\ra, 
\ver 2 2 \ra \}$.
After Alice makes a measurement, the state of Bob's qutrit can
be $\ver \psi_0 \ra$, or $\ver \psi_1 \ra$, or $\ver \psi_2 \ra$.
Since Alice knows the state of Bob's qutrit, she has to convey 
to Bob by classical communication whether to apply $U_{01}$,
or $U_{02}$, or do nothing. This she can do using $\log_2 3$
classical bits.
Thus, the protocol is successful all the times and Alice is able to 
remotely prepare this ensemble of qutrit states with probability one. 
The number of cbits used is $\log_2 3$ and the number of ebits used
is $\log_2 3$ if we consider entanglement between particles $(1,2)$ 
versus 3.

\subsection{ Exact remote state preparation for multiparties}

Next, we consider the case of more than one party. First we 
consider the simpler case when Alice wishes to remotely prepare 
a state at two locations {\it simultaneously}. Let Bob and Charlie
be at these locations. Afterwords, we can generalize to the case
of arbitrary number of parties, say $m$.

In the two locations case, Alice, Bob and Charlie can have one qutrit
each, which are together in a dark state. 
Suppose Alice possesses particle $1$, Bob has particle $2$ and
Charlie has particle $3$. Then after applying $U^{\dagger}$ to her
particle she does a single particle measurement. If
her outcome is $\ver 0 \ra$, then the state at Bob and Charlie collapses to
${1\over \sqrt 2} ( \ver \psi_1 \psi_2 \ra - \ver \psi_2 \psi_1 \ra)$ 
which is an entangled state!
It is not possible to transform it to the desired state 
$\ver \psi_0 \ra$ using LOCC. So RSP cannot be carried out in this way.  
Similar situation occurs when Alice obtains $\ver 1 \ra$ and 
$\ver 2 \ra$ upon measurement. 

So one needs a dark state with more than three particles.
>From the condition of section II, we have $N = 3 m $, where $m$ is a
natural number, denoting the number of locations in our context. 
So we need a minimum of six qutrits to carry out RSP for two parties. 
If Bob and Charlie have more than one of these
qutrits, then as earlier, the RSP protocol cannot be carried out.
This is because their qutrits would be in entangled state after
Alice makes a measurement on her qutrits. So Bob and Charlie
would have one qutrit each, while Alice would have four qutrits.
For  the six qutrits, we can choose a dark state by simply
taking the tensor product of the resource state $(23)$, i.e.,
$\ver \Phi \ra _{123} \otimes \ver \Phi \ra_{456}$
where Alice possesses particles $(1,2,4,5)$ while Bob and Charlie have
particles $3$ and $6$ respectively.
Alice now can remotely prepare a state from the ensemble $(25)$.
To start the protocol, Alice makes unitary transformations on
her qutrits as discussed in the last section. Then she makes
a {\it four-particle measurement} on her qutrits. This she does
in a basis obtained from the tensor products of computational
basis of a qutrit. After her measurement, the state of Bob's and
Charlie's qutrit would be
 $\ver \psi_0 \ra$, or $\ver \psi_1 \ra$, or $\ver \psi_2 \ra$.
Since Alice knows the state of Bob's and Charlie's qutrits,
she has to convey 
to them by classical communication whether to apply $U_{01}$,
or $U_{02}$, or do nothing. This she can do by sending $\log_2 3$
classical bits each to Bob and Charlie. So total information cost
is $2\log_2 3$ cbits and $2 \log_2 3$ ebits.

This can be immediately generalized to the case of $m$ parties.
In such a case, Alice and $m$ parties need to share a dark state
involving $3m$ qutrits. Of these $2m$ qutrits would be with Alice
and one qutrit each with $m$ parties. One example of such a dark
state can be obtained in parallel to two parties case by taking the appropriate
tensor product of the state $(23)$. Alice needs to make a measurement
on her $2m$ qutrits and send $\log_2 3$ cbits to each party to
convey what transformation to apply. One would use $m\log_2 3$ 
cbits and $m \log_2 3$ ebits in the process.

\subsection{ Probabilistic remote state preparation of a qutrit}

We have seen above how the RSP protocol works for the states
belonging to specific ensembles. We also remarked that
the protocol does not work for a general qutrit state.
Here we wish to consider the general qutrit state $(21)$
again and see what is possible.

We examine the simplest situation where Alice wishes to remotely
prepare the state $(21)$ at Bob's location. Alice has two qutrits
and Bob has one. These qutrits are in the dark state $(23)$.
Now suppose Alice applies unitary transformation as given in $(24)$
where $\ver \psi_0 \ra,\ver \psi_1 \ra$and  $\ver \psi_2 \ra$
are as given in $(22)$.  We see that if Alice's result is 
$\ver 1 2\ra$ or $\ver 2 1 \ra $
then Bob's state is $\ver \psi_0 \ra$ which is the desired state
that Alice wants to prepare. For the remaining measurement results, Bob 
would have either $\ver \psi_1 \ra$ or $\ver \psi_2 \ra$.
He would then have to apply some unitary operators to transform
these states to $\ver \psi_0 \ra$. It is extremely difficult
to find such general operators independent of 
the  parameters $\gamma_1, \gamma_2, \delta , \phi$ . 
So the success probability of the protocol is $1/3$. However, this
probability can be enhanced if we fix the value of one of the parameters. 
Therefore, let us set $\gamma_1 = \pi /4$. The states reduce to
\begin{eqnarray}
\ver \psi_0 \ra & = &  {1 \over \sqrt 2}( \ver 0 \ra + cos \gamma_2 
e^{i\delta} \ver 1 \ra + sin \gamma_2 e^{i\phi} \ver 2 \ra ) \nonumber\\
\ver \psi_1 \ra & = &  {1 \over \sqrt 2} (\ver 0 \ra -  
cos \gamma_2 e^{i\delta} \ver 1 \ra -  sin \gamma_2 e^{i\phi} 
\ver 2 \ra  \nonumber\\
\ver \psi_2 \ra & = &  sin \gamma_2 e^{i\delta} \ver 1 \ra
- cos \gamma_2 e^{i\phi} \ver 2 \ra 
\end{eqnarray}

It is easy to see that the unitary operator taking $\ver 
\psi_1 \ra $ to $\ver \psi_0 \ra$ is $U_2 = diag ( 1, -1, -1)$.
Notice that $\ver \psi_2 \ra$ is a two dimensional state
and in order to convert it to $\ver \psi_0 \ra$ ( which is three 
dimensional), the unitary operator would surely involve the
parameters.  We would regards this case as a failure.
Thus by fixing one parameter and leaving the other three free, 
we are able to prepare a qutrit
state remotely with success probability $2/3$.  
The key feature in our protocol is the two-particle measurement performed
by Alice.

\subsection{ Joint remote state preparation for a qutrit}

We close this section by imagining a somewhat different 
scenario. This we illustrate by considering the case
involving Alice, Bob and Charlie each possessing one qutrit. 
We have seen earlier that this situation gives an entangled state
between Bob and Charlie after Alice performs a single particle
measurement and hence not useful for RSP.
Let us suppose that she is allowed to collaborate 
with a second party, say Bob. This means that 
Bob also has complete knowledge about the 
qutrit. Then, Alice and Bob situated at two different locations can 
jointly prepare a qutrit state from the ensemble $(25)$ at a 
remote location say, Charlie. Let us see how this is achieved.

Alice , Bob and Charlie share the entangled three particle
resource state considered earlier.
Alice applies appropriate unitary operator, measures her qutrit and 
conveys her result, say $\ver 0\ra$ 
to both Bob and Charlie. Bob applies another unitary operator, 
makes a measurement on his qutrit 
and sends his result to Charlie.
So if Bob gets $\ver 1 \ra$ , then Charlie gets $\ver \psi_2 \ra$, 
and if he gets $\ver 2 \ra$ then $\ver \psi_1 \ra$ is 
prepared at Charlie's location which is the desired state. 
Whatever Charlie gets, he can always make use of the unitary operators 
$U_{01}$ and $U_{02}$ given earlier to transform his state to 
the desired state with probability one.  
For each classical communication, $\log_2 3$ cbits are used.
Thus the protocol is also successful in this kind of a situation 
where two parties collaborate to remotely prepare a special class of 
qutrit states for a third party. It can also be extended to higher
dimensions and for more number of particles.
This bears similarity with the process of secret sharing \cite{hbb,cgl} which
may be worth exploring in future.

\section{ Remote state preparation of a qudit}

Here, we wish to generalize RSP protocol to systems with 
larger than three-dimensional Hilbert space. So,
Alice wants to prepare a $d$-dimensional quantum state at one or
multiple locations. A general state of a $d$-dimensional 
system can be written as:

\begin{eqnarray}
\ver \psi \ra = \sum_{j=0}^{d-1} \beta_j \ver j \ra
\end{eqnarray}
where
\begin{eqnarray}
& &\beta_0 = cos\gamma_1,  \nonumber\\
& &  \ldots \ldots \ldots \nonumber\\
& & \beta_{d-3} = e^{i\alpha_{d-3}} cos\gamma_{d-2} sin\gamma_{d-3} 
                   sin\gamma_{d-2} \cdots sin\gamma_1, \nonumber\\
& & \beta_{d-2} = e^{i\alpha_{d-2}} cos\gamma_{d-1} sin\gamma_{d-2} 
                   sin\gamma_{d-3} \cdots sin\gamma_1, \nonumber\\
&& \beta_{d-1} = e^{i\alpha_{d-1}} sin\gamma_{d-1} sin\gamma_{d-2} \cdots
                           sin\gamma_{1}
\end{eqnarray}
such that the $2(d-1)$ real parameters have the range
$0 \leq \gamma_1, ..., \gamma_{d-1} \leq \pi/2$ and
$0 \leq \alpha_1, ..., \alpha_{d-1} \leq 2\pi$.

 As earlier, this state cannot be prepared using RSP protocol.
However, by making choices of some parameters, one can have
ensembles of states which can be remotely prepared. We will
give one example of such an ensemble. This ensemble will be
a generalization of $(25)$ for the case of qudits.

\subsection{ Exact RSP of a qudit for one and multiparties}

The RSP protocol requires a dark state as a quantum resource
and an appropriate ensemble of states. Let us first consider
the case of one-party. Alice wishes to prepare a state at
Bob's location.
First of all she needs a resource which is shared with Bob. 
As discussed above, for $d = N$, there exists a totally antisymmetric
$N$ particle quantum state of the form
\begin{eqnarray}
\ver \Psi_N \ra = {1 \over \sqrt {N !}} \sum_{\pi} (-1)^{sgn(\pi)}
\ver \pi_1 \ra ... \ver \pi_N \ra
\end{eqnarray}
where $\{\ver \pi_1 \ra, ...,\ver \pi_N \ra = \ver 0 \ra, ..., \ver N-1 \ra
\}$
denotes the orthonormal basis of the Hilbert space. The sum appearing in
the above expression runs over all possible permutations $\pi$
of the $N$ elementary quantum systems considered. 
Out of $N$ particles, Alice possesses $(N-1)$ of them while Bob has 
only a single particle. Having the above
resource at her disposal, Alice makes a measurement on  
her $(N-1)$ qudits and conveys the result to Bob.
Bob's ability to transform the state at his end to the state
desired by Alice would  depend upon choice of the state.

Next, we need an appropriate ensemble of states.
A generalization of the ``equatorial'' qutrit states can be represented 
by
\begin{eqnarray}
\ver \psi_0 \ra = {1 \over \sqrt d} \sum _{j=0}^{d-1} 
e^{i\alpha_j} \ver j \ra
\end{eqnarray}
where $\alpha_0 = 0$. We can obtain this state from $(30)$ by
appropriate choice of angles.

We can construct the whole set of basis vectors, including
$\ver \psi_0 \ra $, by converting the computational basis into 
the discrete Fourier transform basis as follows:

\begin{eqnarray}
\ver \psi_k \ra = {1 \over \sqrt d} \sum _{j=0}^{d-1} \Gamma^{jk}
e^{i\alpha_j} \ver j \ra
\end{eqnarray}

Here $\Gamma = e^{2 \pi i/d}$ and $\Gamma^{jk} = (\Gamma)^{jk}$
The above set $\{ \ver \psi_k \ra , k= 0,1,...,d-1 \}$ forms an
orthonormal basis. Note that $k = 0$, corresponds to the special
qudit state $(33)$ Alice has chosen to prepare remotely.

Alice first makes unitary transformations on her qudits,
so that in the dark state $(32)$, Bob's qudit can be thought 
of to be in one of the $\ver \psi_k \ra$ states. When
Alice performs a measurement on her $N-1$ particles,
then Bob's qudit will collapse to one of the $\ver \psi_k \ra$
states. Alice then conveys to Bob through $\log_{2} d$
cbits what unitary transformation to apply on his qudit to
transform it into $\ver \psi_0 \ra =\ver \psi \ra$ state.
This is because when Bob gets any one of the basis states 
$\ver \psi_k \ra$, he can apply the corresponding unitary operator

\begin{eqnarray}
U_{0k} = \sum _{j=0}^{d-1} \Gamma^{-kj} \ver j \ra \la j \ver
\end{eqnarray} 
and convert his state to the desired state $\ver \psi_0 \ra$.
Thus the protocol is successful all the times in exactly
preparing a special class of qudit states remotely.

As in the case of qutrits, we can generalize the protocol
for the case of $m$ parties.
In order to prepare a qudit at $m$ locations, we consider an entangled
state consisting of $N = md$ qudits, of which Alice would possess $m(d-1)$  
qudits and the remaining $m$ qudits are distributed among $m$ different 
parties. One can choose, as a quantum resource, a tensor product
of the state $(32)$. Following the usual procedure, Alice projects 
onto her qudits and is able to prepare this special class of qudit 
quantum states remotely at $m$ locations by sending to each party
$\log_2 d$ cbits. This will allow each party to know which one of
the operators given in $(35)$ to apply. If we consider the entanglement
between
Alice's $m(d-1)$ particles verses the remaining $m$ particles, then 
the number of ebits used would be $m\log_2 d$. Total cbits would
be $m\log_2 d$.

\subsection{ Probabilistic RSP of a qudit}

As we know, if Alice attempts to prepare the most general qudit,
 state $(30)$ like in the qutrit case, she would succeed only with 
probability $1/d$. This situation can be improved 
by assigning specific values to some parameters. In one set of
basis vectors, we could have two $d$-dimensional states
and the others of lesser dimensions. Hence we conjecture that 
using this basis, Alice can remotely prepare a fairly general qudit
state with probability $2/d$. For other sets of basis vectors, which
include the state $(30)$, the probability of remote preparation
would be less than $2/d$. This is because of difficulty in finding
parameter-independent unitary transformations connecting various
basis vectors.
 
To gain some insight into the conjecture, let us consider the
$d = 4$ case explicitly.  The most general quantum state for such a
particle can be written as

\begin{eqnarray}
\ver \psi_0 \ra &=& cos\gamma_1 \ver 0 \ra + sin\gamma_1 cos\gamma_2  
e^{i\delta} \ver 1 \ra + sin\gamma_1 sin\gamma_2 cos\gamma_3 e^{i\phi}
\ver 2 \ra 
+ sin\gamma_1 sin\gamma_2 sin\gamma_3 e^{i\sigma} \ver 3 \ra
\end{eqnarray}
Our protocol requires that this should be one of the basis states.
One possible choice for a basis is given by the following
normalized states which are orthogonal to the above state.
\begin{eqnarray}
\ver \psi_1 \ra &=&  sin\gamma_1 \ver 0 \ra - cos\gamma_1 cos\gamma_2
e^{i\delta} \ver 1 \ra - cos\gamma_1 sin\gamma_2 cos\gamma_3 e^{i\phi}
\ver 2 \ra - cos\gamma_1 sin\gamma_2 sin\gamma_3 e^{i\sigma} \ver 3 \ra
\nonumber\\
\ver \psi_2 \ra &=&  sin\gamma_2
e^{i\delta} \ver 1 \ra - cos\gamma_2 cos\gamma_3 e^{i\phi}
\ver 2 \ra - cos\gamma_2 sin\gamma_3 e^{i\sigma} \ver 3 \ra \nonumber\\
\ver \psi_3 \ra &=&  sin\gamma_3 e^{i\phi}
\ver 2 \ra - cos\gamma_3 e^{i\sigma} \ver 3 \ra
\end{eqnarray}  

Analogous to the qutrit case, we fix $\gamma_1 = \pi/4$
while the other five parameters are free.
Following the usual procedure, if Bob gets $\ver \psi_0 \ra$
then he has to do nothing.
If he gets $\ver \psi_1 \ra$
then he applies the unitary operator $U = diag (1, -1, -1, -1 )$
to transform it to the desired state.
When Bob gets the other two states, we shall consider that situation
a failure as those are 2 and 3 dimensional states and the unitary 
operators would always involve the parameters.
Therefore, we are able to achieve RSP of the above special class of
states with probability $1/2$ which is better than a random guess.

\section{Conclusions}

In this paper we have taken a simple approach at generalizing remote state
preparation protocol for special class of states for multiparties. We 
have generalized the protocol for qubits, qutrits and qudits as well. 
The crucial feature of the extension of the protocol is 
the use of multiparticle measurement and the use of dark states 
as a quantum resource. We find that Alice needs to send only $\log_2 d$ cbits
of classical information to each party and consume $\log_2 d$ ebits of 
entanglement per party for remote preparation of a qudit. 
An interesting point to note here is one can prepare special class of states in
any Hilbert space dimension $d$ which does not contradict a result of 
\cite{zeng}. The key observation is that we use different class of 
quantum resource and multiparticle measurements in our protocol. 
However, all dark states are not useful 
for remote state preparation.  In some such cases
only probabilistic remote state preparation for multiparties
is possible.  We hope that this
will provide insight for generalization of remote state preparation for
arbitrary states of qudits at multiparties with low or high entanglement
(asymptotic) limit. 
In the case of qutrits and qudits, we have discussed only one ensemble
of states, for which remote state preparation is possible.
(Of course, the ensembles of states orthogonal to the discussed
ensemble can also be remotely prepared in similar manner.)
There should be many other such ensembles of states. One needs
a systematic procedure to identify such ensembles.
In future, one may also explore how Alice can prepare different Hilbert space
quantum systems at different parties. A tentative line of thought would be
to explore some generalized form of dark states as a quantum resource 
which would be invariant under $U(d_A)\otimes V(d_B) \otimes W(d_C)$, 
with $d_A, d_B, d_C$ being the dimension of three different Hilbert spaces.

\vskip 1cm

{\bf Acknowledgments:} PP acknowledges financial assistance from DST
under the SERC Fast Track Proposal scheme.
AKP thanks A. Thapliyal and D. Leung for useful 
discussions on multiparty RSP during his visit to MSRI, Berkeley, California
during August 26-Sept 26th, 2002. Financial support and hospitality for 
this visit from MSRI is gratefully acknowledged.

\vskip .5cm

\nopagebreak
\renewcommand{\baselinestretch}{1}

\end{document}